\documentclass[%
reprint,
superscriptaddress,
twocolumn,
amsmath,amssymb,
floatfix
]{revtex4-1}
\usepackage{graphicx}% Include figure files
\usepackage{dcolumn}% Align table columns on decimal point
\usepackage{bm}% bold math
\usepackage{enumerate} 
\usepackage{physics}
\usepackage{color}
\usepackage[normalem]{ulem} 
\usepackage{cancel}
\usepackage{soul}
\usepackage{textcomp}

\usepackage{hyperref}% add hypertext capabilities
\begin{document}

\title{Localization-based two-photon wave-function information encoding}

\author{Raffaele Santagati}\email{raffaele.santagati@bristol.ac.uk}\affiliation{Quantum Engineering Technology Labs, H. H. Wills Physics Laboratory and Department of Electrical and Electronic Engineering, University of Bristol, Bristol BS8 1FD, UK}\author{Alasdair B. Price}\affiliation{Quantum Engineering Technology Labs, H. H. Wills Physics Laboratory and Department of Electrical and Electronic Engineering, University of Bristol, Bristol BS8 1FD, UK}\affiliation{Quantum Engineering Centre for Doctoral Training, H. H. Wills Physics Laboratory and Department of Electrical and Electronic Engineering, University of Bristol, Bristol BS8 1FD, UK}\author{John G. Rarity}\affiliation{Quantum Engineering Technology Labs, H. H. Wills Physics Laboratory and Department of Electrical and Electronic Engineering, University of Bristol, Bristol BS8 1FD, UK}\author{Marco Leonetti}\email{marco.leonetti@nanotec.cnr.it}\affiliation{CNR NANOTEC-Institute of Nanotechnology c/o Campus Ecotekne,  University of Salento, Via Monteroni, 73100 Lecce, Italy}\affiliation{Center for Life Nano Science@Sapienza, Istituto Italiano di Tecnologia, Viale Regina Elena, 291 00161 Roma, Italy}

\date{July 22, 2019}% It is always \today, today,

\begin{abstract}
In quantum communications, quantum states are employed for the transmission of information between remote parties. \textcolor{black}{This usually requires sharing knowledge of the measurement bases through a classical public channel in the sifting phase of the protocol.}
 Here, we demonstrate a quantum communication scheme where the information on the bases is shared ``non-classically'', by encoding this information in the same photons used for carrying the data. This enhanced capability is achieved by exploiting the localization of the photonic wave function, observed when the photons are prepared and measured in the same quantum basis. We experimentally implement our scheme by using a multi-mode optical fiber coupled to an adaptive optics setup. We observe a decrease in the error rate for higher dimensionality, indicating an improved resilience against noise.
\end{abstract}
\maketitle

%\tableofcontents

\section{Introduction}

Quantum communication exploits physical properties such as superposition and entanglement to transmit information~\cite{Gisin:2007bya, Krenn:2016jp,Scarani2009SecurityQKD,Pirandola2019}.
Measuring a quantum state collapses the wave function into one of the basis states~\cite{Jones:2011kh, Bassi:2013bp, wheeler2014quantum, Fuwa:2015jv}. In quantum communication, the sender prepares each state on a particular basis, such that when the receiver measures correctly, they can extract the information~\cite{Gisin:2007bya}. If the state is measured on a different basis than the one in which it was prepared, it is not possible to perform a deterministic read-out of the information sent~\cite{Nikolopoulos:2006bh}.
This means that, in a two-party quantum communication protocol, Alice and Bob must exchange the basis information over a classical channel~\cite{Krenn:2016jp}. In simpler terms, qubit measurements are useless without the basis information, and vice-versa. 

Multidimensional quantum communication protocols offer several advantages~\cite{Cerf:2002fp, Bouchard:vo, Durt:2004cu, Nikolopoulos:2006bh} and have been investigated using different technologies, from orbital angular momentum 
\cite{leach2002measuring,walborn2006quantum, MolinaTerriza:2007ig,Zhou:2007bi,DAmbrosio:2012bk, Vallone:2014dj} to path-encoded qudits~\cite{Wang:2018gh, Ding2017highdimensionchipqkd}, enabling the use of higher-order alphabets \cite{pan2017quantum,walborn2006quantum,AliKhan:2007ia}, or to encode multiple qubits in a single photon \cite{tentrup2017transmitting}.
Recent results~\cite{Wang:2018gh,Sibson2017chipqkd,Price2018,Ding2017highdimensionchipqkd} have shown that modern photonics can achieve multidimensional quantum state manipulation, and pave the way for practical applications of quantum communication protocols exploiting multidimensionality.

Here, we introduce an approach for quantum communication, which prepares the photons used for carrying data such that they are transporting both the message and the measurement information in their wave functions. The properties of the wave functions are probed by sending two copies of the same state (i.e. two photons) through the quantum channel. We discuss a potential benefit for quantum key distribution (QKD) and, while we do not examine the security of any particular scheme, note that as a bare minimum the photon pairs would need to be interwoven randomly with each other to avoid the most trivial of intercept-resend attacks. This means the pairings must be publicly announced instead of the bases, to create an information asymmetry between the receiver and any eavesdroppers. As a result, we will need to include a modified sifting phase, rather than being able to remove it altogether as other protocols have managed~\cite{Grazioso2013,Price2017}.

We extend the convention of interweaving the pairs to our generic communications protocol, called multidimensional data basis shuffling (DBS), largely because it brings it closer to being useful for QKD, which is the most dominant form of quantum communication at present. For applications where security is not a concern, this step can be considered optional.

\section{Data basis shuffling}
In the simplest case, we can consider two photons that are transmitted as qubits encoded in the two mutually unbiased bases $\{\ket{0},\ket{1} \}$ or $\{\ket{+},\ket{-} \}$. We call these pairs of photons {\it{twins}}.
\begin{center}
\begin{tabular}{ c c c }
Bit & Basis                    & State sent \\ 
 0  & $\{\ket{0}, \ket{1} \}$  & $\ket{00}$ \\  
 1  & $\{\ket{0}, \ket{1} \}$  &  $\ket{11}$ \\   
 0  & $\{\ket{+}, \ket{-} \}$  &  $\ket{++}$ \\  
 1  & $\{\ket{+}, \ket{-} \}$  &  $\ket{--}$    
\end{tabular}
\end{center}
\begin{figure*}
\centering
\includegraphics[width=0.8\textwidth]{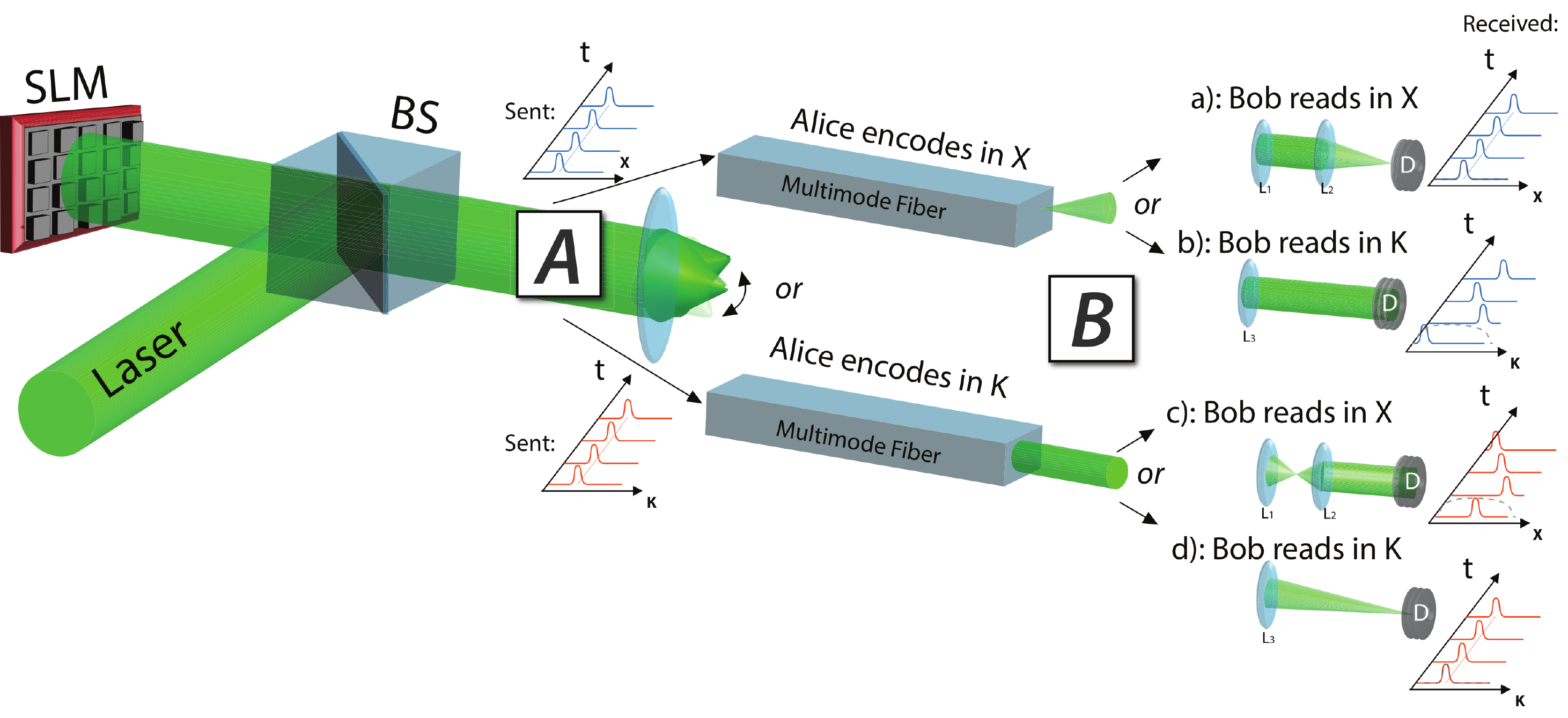}
\caption{Alice sends an attenuated laser into a reflective spatial light modulator (SLM), which focuses the light on the far side of a multi-mode fiber in two mutually unbiased bases. Bob may choose whether to measure in the computational or the Fourier basis by exploiting a flip mirror (not shown). If he measures correctly, the probability density is localized, allowing him to retrieve the bit sent by Alice and confirming his basis choice. If he measures incorrectly, Bob can be expected to obtain inconsistent results as the probability density is delocalized. }\label{fig:Siftless}
\end{figure*}
Each twin is measured with a random projector, which will be one of the following:
\begin{center}
\begin{tabular}{ c c }
$\ket{00}\bra{00}$&$\ket{11}\bra{11}$ \\
$\ket{++}\bra{++}$&$\ket{--}\bra{--}$
\end{tabular}
\end{center}
For example, when sending the state $\ket{00}$, we have two possible scenarios. If the measurement basis is $\{\ket{0},  \ket{1} \}$, the two photons will collapse into $(\ket{00} \bra{00}) \ket{00} = \ket{00}$, returning the result $0$ while also confirming the correctness of the basis because the individual photon states are the same. On the other hand, if the photons are measured in $ \{\ket{+}, \ \ket{-} \}$,
they will have a non-zero probability of collapsing into different states, e.g. $(\ket{++} \bra{++}) \ket{00} = \frac{1}{4}(\ket{00} + \ket{01}+\ket{10}+\ket{11})$. In this case, the result indicates the wrong choice of measurement basis with $50\%$ probability.  {\color{black}We note that, in general, the twins do not need to be sent one after the other, and photon pairs may be interwoven if a much longer string of qubits is sent. This feature means Data Basis Shuffling may potentially be able to act as a foundation for novel QKD protocols, as will be discussed later.}

We now consider a more general case, in which we can encode into bases of higher dimensionality, meaning we are able to transmit information using qudits. Here, the computational basis is described by
\begin{equation}\label{eq:quditcompbase}
\{\ket{k}\}_{k=0, \dots, D-1}=\{\ket{0},\ket{1},\ket{2},\dots, \ket{D-1}\}  ,
\end{equation}
and the Fourier basis by $\{\ket{f}\}_{f= f_0,f_1,\dots,f_{D-1}}$, where the $l^{th}$ Fourier basis state is defined to be 
\begin{equation}\label{eq:quditfouribase}
\ket{f_l} = \frac{1}{\sqrt{D}} \sum_{k=0}^{D-1} e^{i \frac{2 \pi k l}{D}} \ket{k}
\end{equation}
These may be mapped onto the orbital angular momentum of a photon \cite{PhysRevA.88.032305} or its spatial position and reciprocal momentum space on the output facet of a fiber \cite{leonetti2016secure}. When measuring in the wrong basis, the probability of the two photons collapsing into different measurement outcomes will increase as $1-\frac{1}{D}$. Therefore, the probability of Bob falsely believing he is in the correct measurement basis decreases with dimensionality, tending towards zero as the number of dimensions tends to infinity.

In order to break the symmetry between the receiver and an eventual malicious operator, the photon pairs may be interwoven, with Alice publicly revealing the order only after Bob's measurement procedure is complete. {\color{black}It should not be possible to identify the partner of any given photon if each half of a twin is randomly positioned inside a long string of qudits, meaning an eavesdropper will have to resort to guesswork if they wish to pair up the photons ahead of Alice's public announcement. For a string of $N$ qudits, the number of combinations that they must guess between is
\begin{equation}
    C=\prod_{j=2,4,6...}^{N}\frac{j!}{2!\left(j-2\right)!}
\end{equation}
meaning we would not expect the eavesdropper to be able to extract any information from the photons they intercept, given a suitably large value for $N$. As an example, the probability to correctly guess all pairings in a string of $N=100$ photons is $1.21\times10^{-143}$, while the probability to correctly guess all of the bases is $7.89\times10^{-31}$. Thus, the knowledge of the pairings is essential, because the transmitter does not share the basis information during the post-processing phase, meaning the only way of obtaining this is to compare the result of a measurement with that of its ``twin'', whose position is unknown at the time of transmission, to everyone except Alice. Once Bob has been told how the pairs are organized within the string of qudits, he can cross-reference his measurement results and assess the correctness of the basis chosen. If an eavesdropper were able to do this ahead of time, they would be able to identify whether or not they had measured correctly, and forward onto Bob only the photon pairs for which their basis choice matched that used by Alice.

The proposed strategy is similar to sifting in standard BB84 because the pairings are publicly announced after Bob has finished measuring all of the qudits. In our protocol, a correctly-measured subset of the bits that were communicated may be extracted once the photon pairings are known, whereas in BB84, bit-extraction requires Bob to possesses the basis information. If Oscar, a non-malicious third party, were to try and randomly guess the pairings in DBS or the bases in BB84, he would increase the error rate on Bob's final bit string to the point where his presence becomes obvious. The pairing/basis declarations happen after the photons have been measured, breaking the symmetry between the receiver and the eavesdropper, implying that there may be potential for our approach to be used in building new QKD protocols.}

Here, we do not provide an exhaustive security proof of the proposed scheme, however, in section IV we report an initial estimation of the potential benefit of this technique.

\section{Experiment}
{\color{black} It has previously been shown that the degree of localization associated with the probability density of the photon's wave function can be used to implement QKD \cite{leonetti2016secure}. In that context, the term ``localization'' has been borrowed by the condensed matter scientific community, where it is connected with the ``degree of localization'' (i.e  the spatial extension of the wave function) which is a relevant parameter in the phenomenon of Anderson localization \cite{anderson1985question}. When a wave function occupies a small part of the available space, it is said to be localized. Here, we will show how the degree of spatial localization of the  probability density function of the photon enables the receiver to correctly assess whether the measurement has been implemented in the correct basis. In other words, the quantum channel contains both the raw bits that need to be transmitted to the receiver and information on the bases in which they have been encoded.}

\begin{figure}
\centering
\includegraphics[width=0.8\columnwidth]{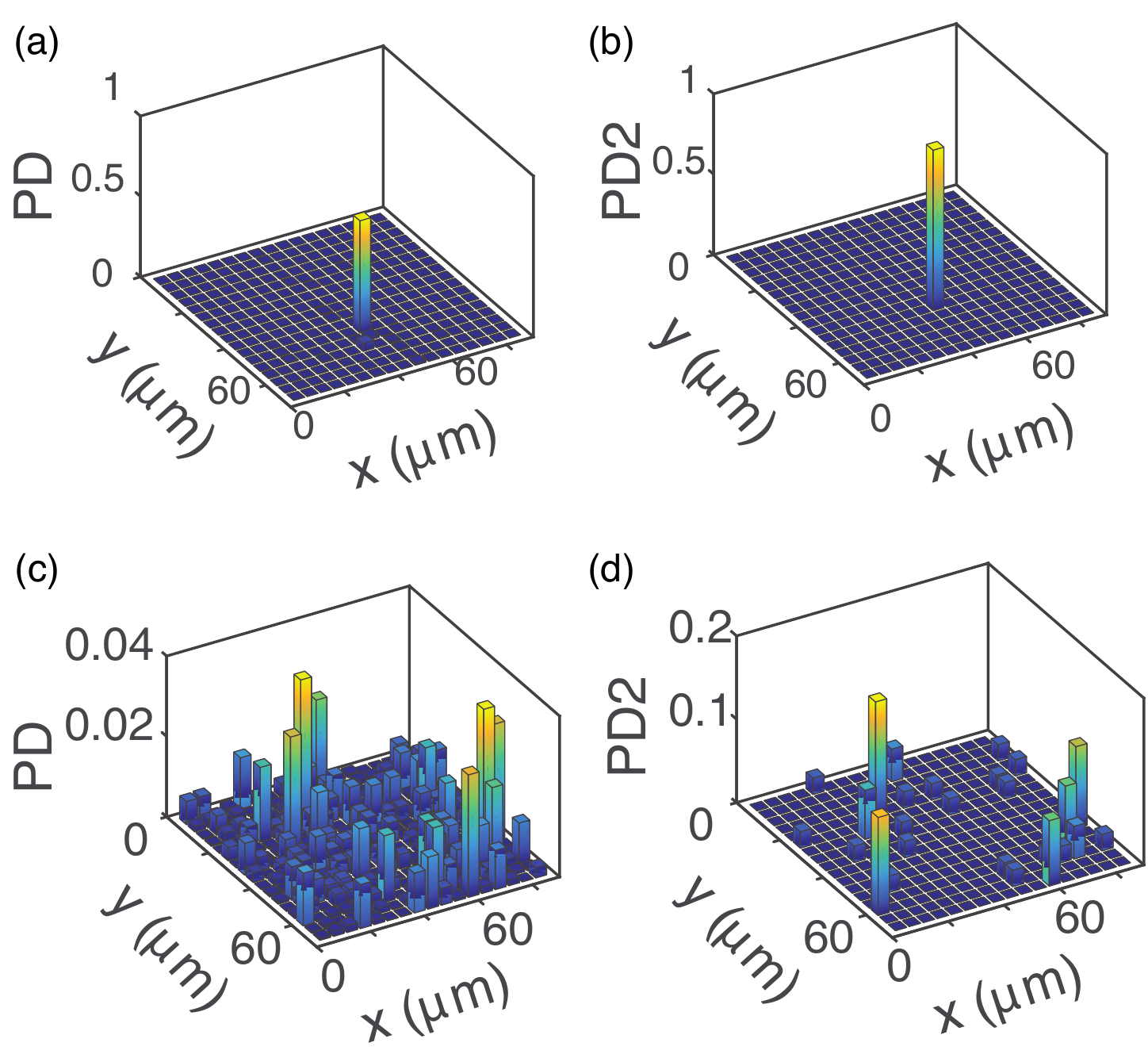}[htp]
\caption{ (a) Probability of detecting only one photon on each pixel of the detector (PD) for a localized state. (b) Probability of detecting two successive single photons on the same pixel (PD2) for a localized state. (c) and (d) respectively PD and PD2 for a delocalized state, which corresponds to a measurement in the wrong basis. All results are for a $17 \times 17$ detector configuration.}
\label{fig: Clicks}
\end{figure}

The experimental scheme of Fig. \ref{fig:Siftless} exploited a multi-mode fiber and a liquid-crystal spatial light modulator (SLM); a promising platform for multidimensional QKD~\cite{Tentrup2018, PhysRevA.94.022315}. In our experiment, the light can be focused at any location in the position space, or in the Fourier basis space \cite{bianchi2012multi,DiLeonardo:11,leonetti2016secure, walborn2006quantum}. 
The SLM is an adaptive optics device with $\mathcal{S}$ segments. A photon reflected by this may be modeled as a scalar plane wave impinging on the $n\text{th}$ segment, whose electromagnetic field is~\cite{Saleh:1991ub,vellekoop2007focusing, PhysRevLett.104.100601}: 
\begin{equation}
E_n=A_n e^{i\phi_n},
\end{equation}
where $A_n$  and $\phi_n$ are the amplitude and phase on the $n\text{th}$ component of a photonic quantum state $\ket{\psi} = \sum_{n=0}^{\mathcal{S}-1} E_n$.
Each $n\text{th}$ input will contribute to the field $E_m$  of the $m\text{th}$ output mode, weighted by the transfer matrix element $t_{nm}$, meaning~\cite{Beenakker1997}
\begin{equation}
E_m=\sum_{n=1}^N |t_{nm}| \sigma_n A_n e^{i \arg\left(t_{nm}\right)}
\end{equation}
where $\sigma_n$ is the state of the $n\text{th}$ SLM pixel (with an arbitrary phase factor in the input mode). We disregard the initial phase $\phi$ as it is a constant term.
Assuming the detector is ideal, the probability to register a photon at a location $\left(x,y\right)$ corresponds to the intensity for the $m\text{th}$ mode,
\begin{align}
\begin{split}
P(x,y)=&E_m^*E_m\\=&\sum_{n=1}^N \sum_{k=1}^N |t_{nm}||t_{km}|  A_n e^{-i \arg\left(t_{nm}\right)}\\&\quad\quad\quad\quad\quad\quad\times A_k e^{i \arg\left(t_{km}\right)}    \sigma_n  \sigma_k.
\end{split}
\end{align}
By choosing the correct $\sigma_n$ and $\sigma_k$ it is possible, through an optimization process~\cite{vellekoop2007focusing,vellekoop2010exploiting}, to obtain a Gaussian focus at an arbitrary location in the $\left(x,y\right)$ plane, which corresponds to the $m\text{th}$ component of a $D$-dimensional qudit. The efficiency of the focusing process is characterized by the signal to noise ratio, often referred to as the enhancement factor. This contributes to the upper bound on the fidelity of the quantum state prepared, and depends also on the number of the segments $\mathcal{S}$~\cite{akbulut2011focusing} of the SLM. The process may be performed for any disordered system described by an arbitrary transmission matrix and in particular for multi-mode optical fibers \cite{bianchi2012multi}.
When a focus is generated in one plane, a delocalized spot is found in the conjugate one, meaning single photons appear in random locations across the area of the detector and repeated measurements are incoherent (see Figs. \ref{fig:Siftless}, \ref{fig: Clicks} and reference~\cite{leonetti2016secure}). {\color{black}This phenomenon connects the degree of localization directly with the capability to assess the correctness of the measurement.}
In our setup, a continuous-wave (CW) laser is modulated by an external trigger, and a variable density filter is positioned before the fiber, attenuating the mean photon number to the desired level. The fiber output is imaged by two lens collection systems with gated single-photon avalanche photodiodes (APDs). For practical reasons, only one detector was used, scanning across the desired spatial positions. By properly adjusting the scanning range, the whole fiber may be imaged on an array with either 16 or 36 detectors, depending on the de-magnification factor of the collection system.
The whole setup is synchronized with respect to a signal generator that controls the laser while also triggering the detectors.
The results of measuring both photons correctly {\color{black}(same basis as Alice)} and incorrectly {\color{black}(different basis to Alice)} are shown in Fig. \ref{fig: Clicks}. It is clear from this that Bob can successfully identify whether or not he chose the right basis. For the ``same basis'' case, Figs. \ref{fig: Clicks}(a) and \ref{fig: Clicks}(b) give the probability to detect a photon (PD) and the probability to detect two successive photons (PD2) in the same detector. Figures \ref{fig: Clicks}(c) and \ref{fig: Clicks}(d) report the same observables for the ``different basis'' case. In the ``same basis'' case, the wave function is localized, so the vast majority of the photon pairs are detected on the same pixel.%If most of photons are detected on the same pixel/position, then the majority of photon pairs are also detected there. 
In the ``different basis'' case, the delocalized wave function gives a non-zero detection probability on all detectors, i.e. a PD with a speckle-pattern-like distribution extending over a large area. It is important to stress that, in this last case, the majority of the pairs (98\%) are measured at different detectors. As a result, these do not contribute to PD2, and are taken into account as ``failed detections''.

\subsection{Additional experimental details}
\label{sec:AppExp}

The silicon avalanche photodiode is a fiber-coupled SPCM-AQR-13 with dark counts of 250~Hz and a dead time of 50~ns. The laser provides $0.5$~$\text{\textmu}\textrm{s}$ pulses with a 2 ns rise time. The spatial light modulator (SLM) is a Holoeye LC-R 768.

Photons are captured using a Galilean telescope, which sets the ratio between the focus size and the detector size, taking care to ensure that the incident light is completely contained within the detector area. 

To measure $\frac{P_{Err}}{P_{Corr}}$, an optimization procedure is performed for each output mode, allowing Alice to store the SLM matrix corresponding to focuses in both position and momentum space. This means $D=16$ or $D=36$ optimizations must be performed for each basis (i.e. 32 or 64 optimizations in total). Once Alice has chosen a letter and a basis, she can select the matching SLM mask from her database and send two photons. Bob measures using a multi-detector array, randomly choosing to monitor either the position or the momentum basis, and keeps note of the detector that clicked. From this, it is possible to extract the probability of each detector clicking for an individual photon. After the set of measurements is complete, and Alice has announced how the photons have been shuffled, the probability of detection can be evaluated for each pair. Bob must check whether the detection events observed match Alice's announcement, to identify when the correct bases were chosen. \textcolor{black}{ In table~\ref{tab:experiment} we report the main parameters observed in this experimental demonstration.}

\begin{table}[h]
    \centering
    \caption{\textbf{Experimental conditions. \boldmath$\eta$ is the Efficiency of the quantum communication channel, $\mathbf{\gamma}$ is the Dark count rate and $\mathbf{\lambda}$ is the Mean photon number.}}
    \label{tab:experiment}
    \begin{tabular}{|c|c|c|}
    \hline
        \boldmath$\eta$  & \boldmath$\gamma$                 &  \boldmath$\lambda$  \\
        \hline
        $0.52$           & $300$ $\left[\textrm{Hz}\right]$  & $0.2$  \\
        \hline
    \end{tabular}
\end{table}

\section{Quantifying errors}
One of the key metrics in quantifying the performance of our quantum communications protocol and comparing it to those that rely on standard individual-photon bit encoding (IPBE), is the ratio between the probability of obtaining an error and the probability of obtaining correct information:
\begin{equation}\label{eq:metric}
\frac{P_{Err}}{P_{Corr}}= \frac{ P_{BE}+ P_{EE}}{ P_{Corr}},
\end{equation}
where $P_{BE}$ is the error probability for measurements in the wrong basis and $ P_{EE}$ is the error probability due to a dark-counts-induced false positive. $P_{Corr}$ is the probability of obtaining the correct information.
In section \ref{sec:AppQuant}, we show that

\begin{equation}
P_{Corr} = \frac{\eta^2}{4}(1-e^{-\lambda})^2\exp^{-2\gamma \tau (D-1)},
\end{equation}
\begin{equation}
P_{BE} =\frac{\eta^2}{4D}(1-e^{-\lambda})^2,
\end{equation} 
\begin{equation}\begin{aligned}
P_{EE} =& e^{-2\lambda} P_{\gamma}^2/D+(1-e^{-\lambda} )^2(1-\eta)^2P_{\gamma}^2/D,
\end{aligned}
\end{equation}
where $\eta$ is the total efficiency of the transmission channel and detectors, $\lambda$ is the mean photon number,  $\gamma$ is the number of dark counts per second and $\tau$ is the inverse of the gate rate
 so that 
\begin{equation}\begin{aligned}
\frac{P_{Err}}{P_{Corr}}\propto \exp^{2\gamma \tau (D-1)}/D.
\end{aligned}
\end{equation}
In the case of IPBE, we have
\begin{equation}
P^{IPBE}_{EE} = e^{-\lambda} P_{\gamma}+(1-e^{-\lambda} )(1-\eta)P_{\gamma}.
\end{equation}

\begin{figure*}
\centering
  \includegraphics[width = 0.8\textwidth]{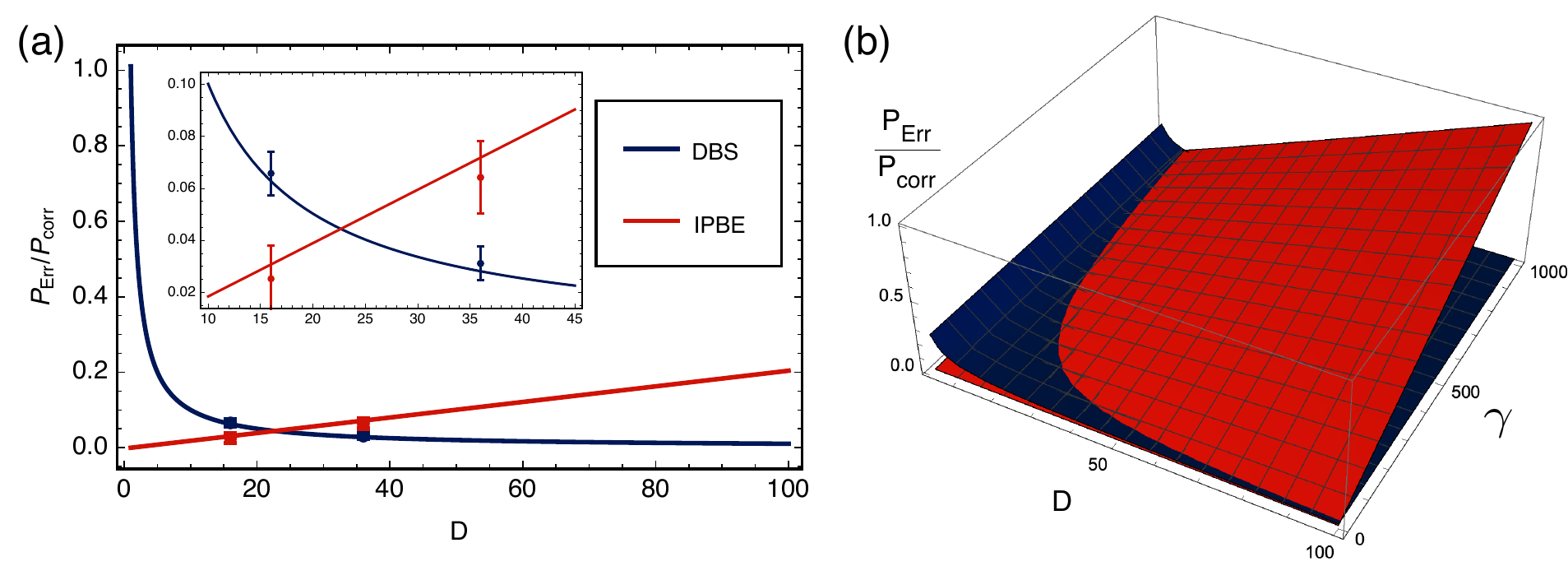}
  \caption{(a) Theoretical and experimental results for the error probability over the probability of successful communication ($P_{Err}/P_{Corr}$) vs the dimensionality ($D$). The inset graph is a magnified view of the region with experimental points. (b) Error probability over the probability of successful communication vs dark counts ($\gamma$) and the dimensionality. While individual-photon bit encoding (red) is more effective for low dimensions, data basis shuffling (blue) quickly becomes less prone to error when the dimensionality is greater than 20. The number of dark counts does not significantly affect this performance metric.}\label{ComparisonExp}
\end{figure*}

\begin{figure*}
\centering
  \includegraphics[width = 0.8\textwidth]{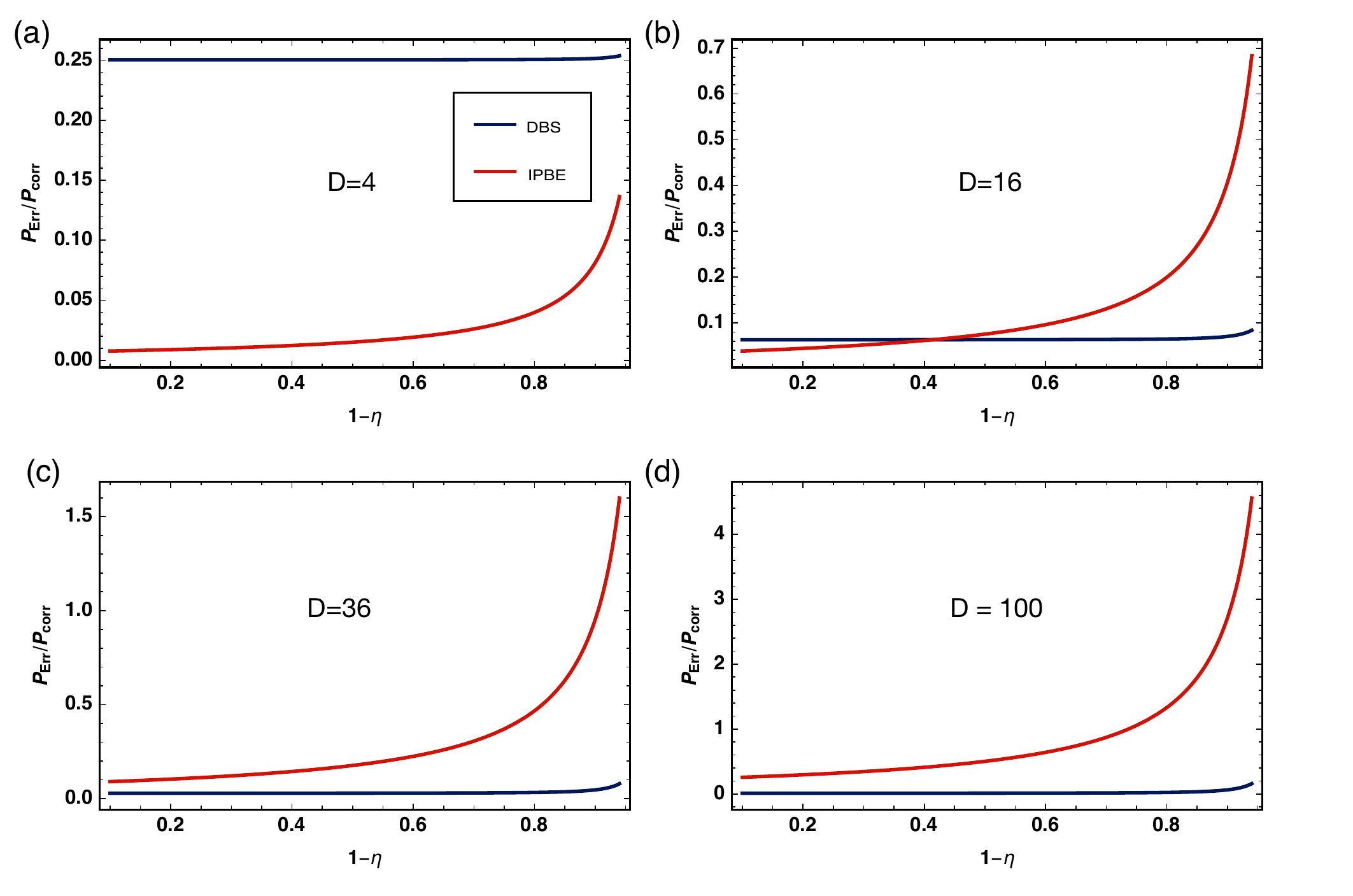}
  \caption{\textcolor{black}{$P_{Err}/P_{Corr}$ as a function of loss ($1-\eta$) for four different dimensionality ($D$) values. $\gamma$ and $\lambda$ are fixed at $500$ dark counts per second and $0.2$ respectively. (a) At $D=4$, IPBE always outperforms DBS, regardless of the loss. (b) At $D=16$, DBS starts to become preferable when the loss exceeds $45\%$. (c) and (d) At higher values of $D$, DBS always outperforms IPBE, regardless of the loss.}}\label{LOSSES}
\end{figure*}

In Fig.~\ref{ComparisonExp}(a), we present an experimental and theoretical comparison of the DBS and IPBE performances, exploiting experimentally retrieved values for $\gamma$, $\eta$ and $\tau$. In Fig.~\ref{ComparisonExp}(b), we report the theoretical three-dimensional plot of $\frac{P_{Err}}{P_{Corr}}$ as a function of $D$ and $\gamma$, showing that DBS outperforms IPBE in an extended area of the $D-\gamma$ space. The extension of this area is controlled by the dark count rate  $\gamma \tau$ (see equation~\ref{eq:Err_Corr}). In our experiments, we used a commercial avalanche photo-diode, providing $\gamma \tau=250$ Hz. However, this value may be improved considerably by introducing a superconducting nano-wire single-photon detector, which can be expected to have a higher efficiency and lower dark counts~\cite{Sprengers2011}. Further improvements can be achieved by using on-chip photon-pair generation, together with state preparation and coincidence-counting logic~\cite{Santagati:2017uf}.

\textcolor{black}{The channel loss can also be detrimental to quantum communications, meaning that assessing the effect of this on DBS and IPBE is important for understanding their performance in realistic scenarios. In Fig.~\ref{LOSSES}, we report the effect of loss ($1-\eta$, where $\eta$ is the channel efficiency) on the $\frac{P_{Err}}{P_{Corr}}$ ratio for $D=4$ in Fig.~\ref{LOSSES}(a), $D=16$ in Fig.~\ref{LOSSES}(b), $D=36$ in Fig.~\ref{LOSSES}(c) and $D=100$ in Fig.~\ref{LOSSES}(d), with fixed values of $\gamma=500$ counts/s and $\lambda=0.2$. While IPBE offers a clear advantage at low $D$ and low loss, DBS is more tolerant to higher attenuation and gives a better performance as $D$ increases.
}

\textcolor{black}{
While the effect of loss on DBS is quadratic, because we use two photons per bit, our scheme can offer an advantage in specific circumstances. Consider an experiment in which $P_{err}/P_{corr}$ is required to be lower than $0.4$, with a total attenuation of $(1-\eta)=50\%$ and $\gamma=450$ counts/s. DBS would communicate information at 0.25 times the rate of IPBE, due to the bit-rate reduction associated with ``twins'' carrying the information instead of single photons, and the fact that only half of a ``twin'' has to go missing for the information to be irretrievable. On the other hand, if the number of dark counts is high ($\gamma >450$ counts/s) then DBS maintains a good performance up to $D=90$, whereas IPBE cannot surpass $D=20$ while staying within the required error rate. In a lossless scenario, each DBS ``twin'' will communicate $4.5$ times the total number of bits compared to a single photon in IPBE, because photons encoded in a higher dimensionality transport more information. Therefore, we can see that using DBS increases the overall bit-rate by a factor of 1.125 relative to IPBE.
 }

\subsection{Calculations of errors}
\label{sec:AppQuant}
In this work, we use the convention that when there is no superscript, e.g. $P_{Corr}$, we refer to the DBS case, while when referring to the IPBE case, we explicitly indicate it in the superscript, e.g. $P^{IPBE}_{Corr}$.

Now, we want to estimate the probability of a result being acknowledged by Bob as a correct bit, when actually he measured on the wrong basis. This is of importance because it affects the coherence between the key generated by Alice and the one retrieved by Bob.
In DBS, two terms are contributing to the total error probability 
$P_{BE}$. If, for the time being, we disregard dark counts,
\begin{equation}
P_{BE} = P_{bothE} + P_{bSinglE},
\end{equation}
where $P_{bothE}$ takes into account the cases when
Bob retrieves both photons in the wrong basis (25\% probability) and $P_{bSinglE}$  indicates when Bob retrieves one of the photons in the correct basis and the other in the wrong one (50\% probability).

However, if Bob chooses two different bases for a pair of photons, both Bob and Alice can discard the retrieved bits from that photon as requested by the DBS protocol (this case is equivalent to IPBE ``wrong basis'' case, in which both Alice and Bob discard the photons).
So the  relevant error is
\begin{equation}
P_{BE} = P_{bothE} =\frac{\eta^2}{4D}(1-e^{-\lambda})^2.
\end{equation}
In IPBE the corresponding  term is always 0 as measurements in the wrong basis will always be eliminated through classical communications.

Now, let us consider the impact of dark counts, which is relevant especially for empty pulses. The probability of obtaining a false click in the presence of dark counts~\cite{Scarani2009SecurityQKD} is: 
\begin{equation}
\begin{split}
 P_{\gamma}
 = 1-e^{-\gamma \tau (D-1)}.
\end{split}
\end{equation}

An empty pulse has a certain probability of being detected as a loaded pulse due to dark counts, increasing the quantum bit error rate (QBER)~\cite{Bouzid2011darkqber}. In DBS, dark counts must replace both photons for this to happen, so the probability becomes
\begin{equation}
P_{EE} = e^{-2\lambda} P_{\gamma}^2/D+(1-e^{-\lambda} )^2(1-\eta)^2P_{\gamma}^2/D.
\end{equation}
In contrast, there is no way to check in IPBE if a message is due to dark counts or light, so the error probability grows exponentially with the dimensionality of the system

\begin{equation}
P^{IPBE}_{EE} = e^{-\lambda} P_{\gamma}+(1-e^{-\lambda} )(1-\eta)P_{\gamma}.
\end{equation}
In the case of DBS, the probability of transmitting correct information is
\begin{equation}
P_{Corr} = \frac{\eta^2(1-e^{-\lambda})^2 (1-P_\gamma)^2 }{4}.\label{eq:PEEIPBE}
\end{equation}
In Eq.~\ref{eq:PEEIPBE}, $\eta^2$ is the probability of both photons to be measured, $(1-e^{-\lambda})^2$ is the probability of having two photons in the transmission channel and $(1/2)^2$ is the probability of measuring them in the correct basis, which is $1/2$ for each of them.

Similarly, for IPBE
\begin{equation}
P^{IPBE}_{Correct} = \frac{\eta(1-e^{-\lambda}) (1-P_\gamma)}{2}.
\end{equation}

As figure of merit, we use the ratio:
\begin{equation}
\frac{P_{Err}}{P_{Corr}}= \frac{ P_{BE}+ P_{EE}}{P_{Corr}}.\label{eq:Err_Corr}
\end{equation}

\begin{figure*}
  \includegraphics[width=\textwidth]{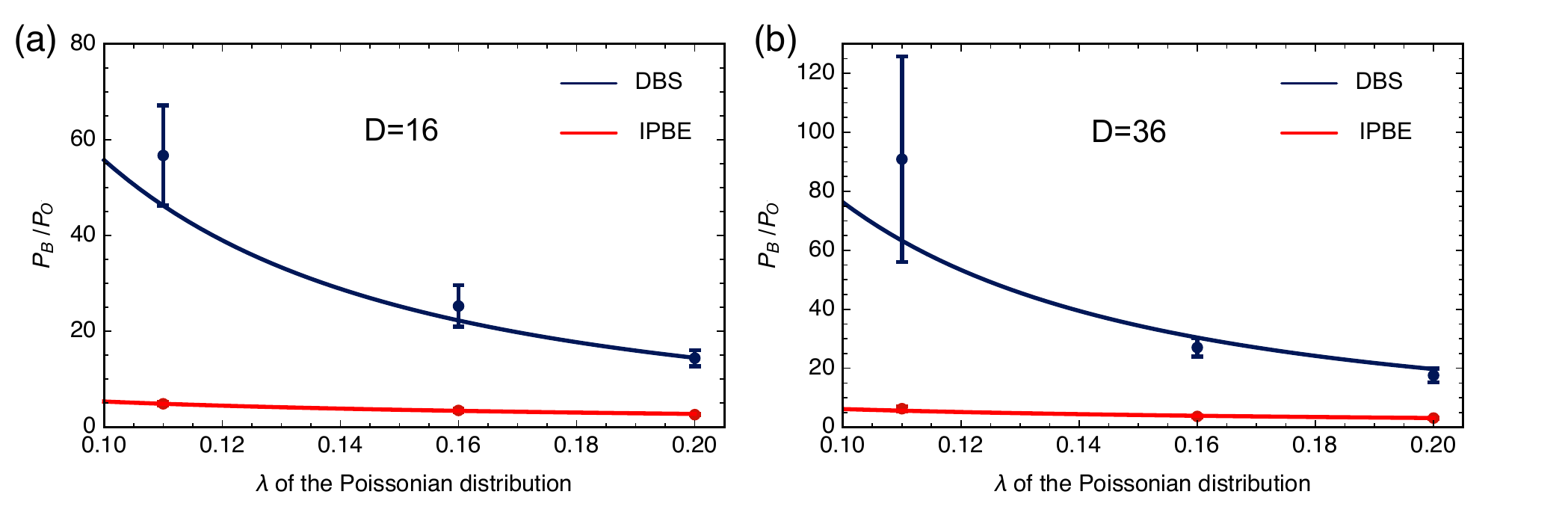}
  \caption{Experimental data points and theoretical curves showing the potential resilience of a DBS-based quantum key distribution protocol to photon number splitting attacks (a) when $D=16$ and (b) when $D=36$.  }\label{ExpResHor}
\end{figure*}

\section{Towards quantum key distribution with localized wave functions}

We now introduce a non-malicious opponent, Oscar, whom we challenge to obtain more information than Bob by splitting multi-photon pulses~\cite{Brassard2000pns}. We look at the success probabilities for Bob and Oscar to extract any information from the photons sent by Alice, in order to give a preliminary estimate for the expected performance of data basis shuffling (DBS).

In the case of Bob, the chances of him being able to extract information successfully will depend directly on the dimensionality of the qudits, the dark counts of the detectors and whether or not multi-photon terms are present. 
Meanwhile, Oscar's probability will mostly depend on the presence of multi-photon terms, assuming he has perfect equipment.

One important quantity for both DBS and individual-photon bit encoding (IPBE) is the probability of obtaining a false click in the presence of dark counts~\cite{Scarani2009SecurityQKD}:
\begin{equation}
\begin{split}
 P_{\gamma}
 = 1-e^{-\gamma \tau (D-1)}.
\end{split}
\end{equation}

Whenever a gate contains at least one photon sent by Alice, the probability of Bob successfully extracting information using DBS is

\begin{equation}
P_B = \left(\frac{\eta}{2}  (1 - P_{\gamma}) \right)^2=  \frac{\eta^2}{4}\exp^{-2\gamma \tau (D-1)},
\end{equation}
where $\eta$ is the combined efficiency of the transmission channel and detectors. The factor of 1/2 arises from having two possible measurement bases.

If we now assume that Oscar has perfect detectors and is restricted to performing only photon number splitting, then the probability that he successfully extracts the same information is

\begin{equation}
P_O = \frac{1}{2} {\left( P_{\textrm{mult}}/P_{\textrm{phot}} \right)}^2,
\end{equation}
where $P_{mult}= \sum_{n=2}^{\infty} \frac{\lambda^n}{n!}e^{-\lambda} $ is the multi-photon probability and $P_{\textrm{phot}} = \sum^{\infty}_{n=1} \frac{\lambda^{n} e^{-\lambda}}{n!}$, is the probability of having at least one photon.

The quantity we use as a benchmark, in this case, is the ratio of the probabilities for Bob and Oscar:

\begin{equation}
\frac{P_B}{P_O} = \frac{\left(\frac{\eta}{2}e^{-\gamma \tau (D-1)} \right)^2}{\frac{1}{2} (P_\textrm{mult}/P_\textrm{phot})^2}.
\end{equation}
In Fig. \ref{ExpResHor}, we compare this quantity with the equivalent in the case of IPBE:

\begin{equation}
\frac{P^{\textrm{IPBE}}_B}{P^{\textrm{IPBE}}_O} = \frac{\frac{\eta}{2}e^{-\gamma \tau (D-1)}}{(P_\textrm{mult}/P_\textrm{phot})}.
\end{equation}

In principle, if Alice and Bob's reference frames are slowly rotating relative to one another, it may be possible to use a non-interwoven variant of DBS to perform direct quantum communication, as identical measurement results from both halves of a twin would indicate the reference frames were aligned. However, this would lead to a substantial reduction in the bit rate because, while DBS would allow Alice and Bob to identify any bits that may be considered unreliable as a result of reference frame misalignment, it does not provide a method for communicating during these periods, which will constitute the majority of the time. In a similar vein, one may consider whether DBS could be used to identify the set of bases being used by Alice, without her announcing it classically before the protocol begins. However, this would be expected to fail, as the probability for Alice and Bob to measure in the same basis, chosen from a virtually infinite set of possibilities, would be close to zero, and far less efficient than any classical method. Thus, the use of DBS as a technique for exchanging information without sharing a previously-defined set of bases seems unlikely to be of any practical use.

\section{Discussion and conclusions}
We have introduced and experimentally demonstrated a new quantum communication protocol, which exploits multidimensional quantum channels.
Our experimental observations, together with the numerical predictions presented herein, are evidence that this protocol can offer improvements over standard individual-photon bit encoding (IPBE) when increasing the dimensionality of the quantum channel. In data basis shuffling (DBS), two copies of the same quantum states are transmitted over the multidimensional quantum channel, allowing Bob to infer the basis information from his measurement results, instead of having to receive it over a classical link. This shows potential for increasing the resilience of quantum key distribution to photon number splitting attacks if it were to be used as the foundation for novel cryptographic protocols. 

At the same time, DBS naturally ``double-checks'' the information received, and the probability of random noise being misinterpreted as a legitimate detection is lower than for IPBE. As a result, the probability of error goes down, implying a higher tolerance towards noise. This effect is especially useful for high dimensionality where the number of detectors goes up, leading to increased dark counts.  

Future studies should focus on characterizing the DBS bit rate for general amounts of loss, exploring how the dimensionality and dark counts affect the resulting trend. From this, it will be possible to perform a comparison with IPBE to identify a complete set of experimental conditions for which DBS outperforms traditional approaches.

\section*{Funding}
Engineering and Physical Sciences Research Council (EPSRC) (EP/L015730/1, EP/M024458/1);   Fondazione CON IL SUD (2015-pdr-0253). 

\section*{Acknowledgments} We thank Phil Sibson, Jake Kennard, Jorge Barreto, Anthony Laing and Paul Skrzypczyk for useful discussions.
\section*{Disclosures} The authors declare that there are no conflicts of interest related to this paper.

\textbf{Data Availability:}
All data used to achieve the conclusions of this work are available for download at \href{url}{https://doi.org/10.6084/m9.figshare.8036105.v1}.

%\bibliographyfullrefs{main}

%=============================================================================

%
%=========================================================

\bibliography{apssamp}% Produces the bibliography via BibTeX.

\end{document}